\documentclass[sigconf]{acmart}
\AtBeginDocument{%
  }

\usepackage{microtype}
\usepackage{graphicx}
\usepackage{subcaption}
\usepackage{booktabs} 
\usepackage{hyperref}
\usepackage{adjustbox}
\usepackage{tabularx}
\usepackage{enumitem}

\usepackage{amsmath}
\usepackage{mathtools}
\usepackage{amsthm}
\usepackage{algorithm}
\usepackage{algpseudocode} 

\usepackage[capitalize,noabbrev]{cleveref}

\theoremstyle{plain}

\theoremstyle{definition}

\theoremstyle{remark}

\newcommand{\systemname}{DynaExq}

\begin{document}

\title{Dynamic Expert Quantization for Scalable Mixture-of-Experts Inference}

\author{Kexin Chu}
\affiliation{
  \institution{University of Connecticut}
  \city{Storrs}
  \state{CT}
  \country{USA}
}

\author{Dawei Xiang}
\affiliation{
  \institution{University of Connecticut}
  \city{Storrs}
  \state{CT}
  \country{USA}
}

\author{Zixu Shen}
\affiliation{
  \institution{University of Connecticut}
  \city{Storrs}
  \state{CT}
  \country{USA}
}

\author{Yiwei Yang}
\affiliation{
  \institution{University of California, Santa Cruz}
  \city{Santa Cruz}
  \state{CA}
  \country{USA}
}

\author{Zecheng Lin}
\affiliation{
  \institution{Independent Researcher}
  \country{China}
}

\author{Wei Zhang}
\authornote{Corresponding author. Email: \texttt{wei.13.zhang@uconn.edu}}
\affiliation{
  \institution{University of Connecticut}
  \city{Storrs}
  \state{CT}
  \country{USA}
}

\author{}

\begin{abstract}
Mixture-of-Experts (MoE) has become a practical architecture for scaling LLM capacity while keeping per-token compute modest, but deploying MoE models on a single, memory-limited GPU remains difficult because expert weights dominate the HBM footprint. Existing expert offloading and prefetching systems reduce the resident set, yet they often pay expert-loading costs on the critical path when activation becomes dense. Post-training quantization (PTQ) lowers the footprint without transfers, but prevailing pipelines fix expert bit-widths offline and assume routing remains stable, even though MoE expert utilization is heavy-tailed and the hot set can shift across workloads.

We present \systemname, a runtime-aware mixed-precision serving system that treats single-GPU MoE inference under a hard HBM envelope as an online, budget-constrained precision allocation problem. The key insight is to keep the experts that dominate runtime traffic resident at higher precision, while maintaining a low-precision fallback for the remaining experts, so the system can reduce transfer volume and avoid the waiting latency that limits offloading and prefetching under dense activation. \systemname\ estimates long-horizon expert hotness from router traces, selects a per-layer high-precision resident set via a budget-feasible top-$n$ rule, and applies promotions and demotions asynchronously through stable expert handles so the forward pass always executes on a fully materialized expert version. Across Qwen3-MoE-30B/80B and six benchmarks, \systemname\ improves accuracy over static PTQ on Qwen3-80B (73.09\% $\rightarrow$ 77.57\%) under comparable device-memory budgets and achieves up to $2.73\times$ higher throughput than offloading/prefetch baselines at batch size 32.
\end{abstract}



\maketitle

\section{Introduction}

The rapid expansion of large language models (LLMs) has driven their adoption across a broad range of applications~\cite{zhao2023survey}. Beyond cloud deployment in centralized data centers, there is growing interest in running LLMs closer to end users to reduce interactive latency, limit exposure of sensitive data, and avoid dependence on stable network connectivity~\cite{friha2024llm}. These pressures have made on-device and edge inference an increasingly relevant deployment mode, but they also expose a practical constraint: the memory footprint of modern LLMs often exceeds the capacity of a single commodity accelerator, especially on edge platforms.

MoE architectures~\cite{zhou2022mixture} have emerged as a pragmatic path to scaling model capacity without incurring proportional per-token compute. By routing each token to a small subset of experts, MoE increases representational capacity while keeping the activated parameters per token relatively small. The reduction in activated compute, however, does not translate into a proportional reduction in storage demand at inference time. To preserve unconstrained routing, the system must keep the full set of expert weights accessible at low latency, which shifts the bottleneck from arithmetic throughput to parameter residency. As a result, MoE-based LLMs can remain memory-intensive even when only a small fraction of parameters is used per token. For example, Qwen3-Next-80B activates only about 3B parameters per token, yet storing the full model parameters requires on the order of 160\,GB of memory, far beyond what typical edge-class GPUs can provide. This gap makes single-device deployment difficult and motivates systems support that treats expert parameters as the primary constrained resource.

Two classes of techniques are commonly used to relax this memory bottleneck. The first is expert offloading and prefetching~\cite{li2023adaptive,yi2023edgemoe, xue2024moe,eliseev2023fast,hwang2024pre,zhong2024adapmoe,tang2024hobbit,song2024promoe,shen2025expertflow}, which treats GPU memory as a cache and moves experts between the GPU and slower tiers such as host memory or SSD. The second is post-training quantization (PTQ)~\cite{lin2024awq,frantar2022gptq,kim2023mixturequantizedexpertsmoqe,duanmu2025mxmoe,fu2025eaquantenhancingposttrainingquantization,frantar2023qmoepracticalsub1bitcompression,zhang2025moqe,hu2025moequantenhancingquantizationmixtureofexperts}, which compresses expert weights into low bit-width representations. Both are effective in specific regimes, but both rely on assumptions that can be violated under realistic serving mixes.

Offloading and prefetching are most effective when each iteration touches a small, stable working set of experts. In practice, MoE activation can become substantially denser during prefill and at larger batch sizes, which expands the per-iteration working set and increases transfer pressure (\autoref{tab:expert_activation_ratio_prefill}). When the working set grows beyond what can be staged over PCIe or NVLink within the available overlap window, transfers become visible as GPU waiting time and amplify tail latency (\autoref{fig:waiting-vs-prompt}). In such regimes, placement policies face a structural limitation: even with accurate prediction, the system must still move a large volume of expert weights to sustain throughput.

PTQ reduces expert footprint without introducing transfer dependencies on the critical path. However, most MoE PTQ pipelines assign a uniform bit-width to all experts or fix a per-expert precision map offline using limited calibration. This design implicitly assumes that the relative importance of experts remains stable at serving time. That assumption is brittle for MoE. Expert utilization is heavy-tailed over long horizons and the identity of frequently used experts can shift across workloads such as general text, math reasoning, and code generation. Under workload shift, a static precision map misallocates scarce high-precision capacity: it either preserves precision for experts that contribute little traffic in the current workload, or over-compresses experts that become hot, degrading quality precisely when the workload is harder.

To address this deployment gap, we propose \systemname, which treats single-GPU MoE serving under a hard HBM envelope as an \emph{online, budget-constrained precision allocation} problem. The key idea is to make expert precision a runtime-controlled resource, rather than a fixed offline choice. \systemname\ continuously observes router outputs during serving, estimates which experts account for a disproportionate share of traffic over a stable time horizon, and allocates a limited high-precision budget to those experts while keeping the remainder in a lower-precision representation to stay within the device memory cap.

\systemname\ is organized as a lightweight control loop that separates policy decisions from the token critical path. On the policy side, a scheduler aggregates routing traces and periodically computes a budget-feasible per-layer resident set for high-precision experts. This selection is constrained by a fixed HBM budget that accounts for non-expert parameters and runtime allocations such as the KV cache, so the resulting precision plan is feasible by construction. On the mechanism side, \systemname\ enforces non-blocking execution by decoupling precision transitions from computation. It maintains stable expert handles that always resolve to a fully materialized expert version, and it performs promotions and demotions asynchronously on a dedicated migration stream. During a transition, the forward pass continues to execute using the last published version of each expert, and version updates become visible only after the corresponding data movement completes. To keep switching predictable under load, \systemname\ also uses deterministic memory partitions for expert weights and explicit admission control for promotions, which prevents allocator fragmentation and avoids transient out-of-memory failures while the serving workload evolves.

The key contributions of this work are:
\begin{itemize}[leftmargin=1.5em, itemsep=1pt, topsep=2pt]
  \item We cast single-GPU MoE serving under a hard HBM envelope as an online, budget-constrained precision allocation.
  \item We design a budget-feasible scheduler that tracks long-horizon expert hotness and selects per-layer high-precision experts via a stable top-$n$ rule.
  \item We build a non-blocking mechanism for mixed-precision residency that decouples precision transitions from the token critical path using stable expert handles and asynchronous promotions/demotions.
  \item We implement \systemname\ and compare it with state-of-art offloading/prefetch and static PTQ method, it improves accuracy over static quantization on Qwen3-80B (73.09\% $\rightarrow$ 77.57\%) and achieves up to $2.73\times$ higher throughput than offloading/prefetch baselines at batch size 32.
\end{itemize}
\section{Background and Motivation}

\subsection{MoE Inference}
MoE layers increase model capacity by routing each token to a small subset of expert subnetworks, typically top-$k$ experts per token. While sparse activation reduces \emph{compute}, it does not directly reduce the \emph{memory} footprint at inference time: to preserve unconstrained routing, a deployment must keep expert weights accessible with low latency. In modern MoE LLMs, expert weights dominate the parameter footprint, so the ability to serve on a single GPU is often determined by whether expert weights can fit within HBM.

A key property of MoE inference is that expert utilization is inherently uneven. Even within a fixed layer, a small subset of experts tends to receive a disproportionate fraction of routed tokens, while many experts are rarely activated. This skew is observable from router outputs alone and does not rely on labels or ground-truth quality measurements. In practice, skew implies that allocating equal fidelity and memory to every expert is rarely cost-effective under a tight HBM envelope.

\subsection{Offloading and Prefetching for MoE Serving}
\label{sec:background-offload}
A widely used approach for fitting MoE inference into limited GPU memory is to treat expert weights as a managed working set. A line of specialized systems exploits sparse expert activation and proposes gating-aware caching and prefetching policies, including Adap-gating~\cite{li2023adaptive}, EdgeMoE~\cite{yi2023edgemoe}, MoE-Infinity~\cite{xue2024moe}, Mixtral-offloading~\cite{eliseev2023fast}, Pre-gated MoE~\cite{hwang2024pre}, AdaMoE~\cite{zhong2024adapmoe}, Hobbit~\cite{tang2024hobbit}, ProMoE~\cite{song2024promoe} and ExpertFlow~\cite{shen2025expertflow}. These methods differ in how they predict upcoming experts and stage transfers, but share a common goal: keep only a subset of experts resident on the GPU and fetch others on demand. Under realistic serving conditions, they may still incur accuracy degradation or latency overhead.

\begin{figure}[t]
    \centering
    \includegraphics[width=0.99\columnwidth]{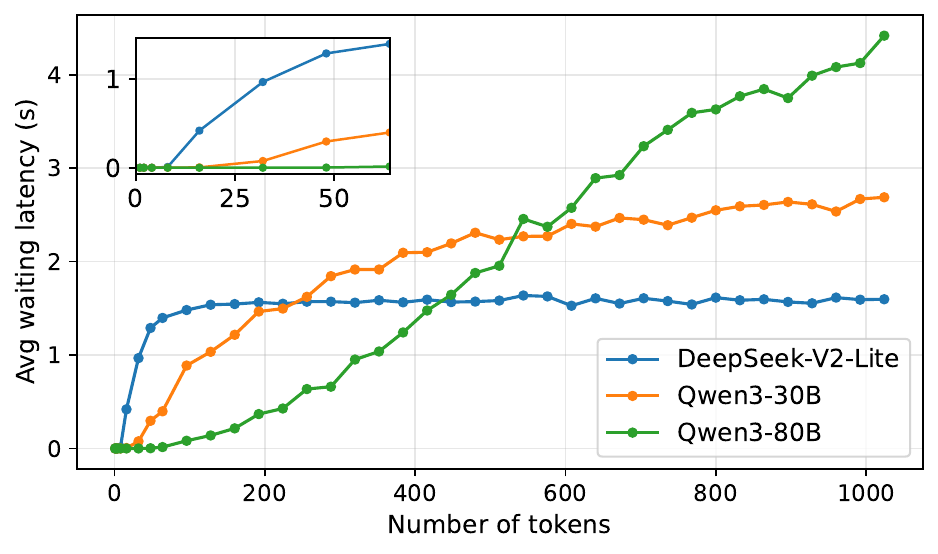}
    \caption{GPU waiting latency vs. the number of tokens.}
    \label{fig:waiting-vs-prompt}
    \vspace{-1.5em}
\end{figure}

Offloading and prefetching are most effective when each layer activates a small and stable subset of experts. Tables~\ref{tab:expert_activation_ratio_decode} and~\ref{tab:expert_activation_ratio_prefill} show that this sparsity regime often does not hold in serving. As batch size increases, the activated-expert ratio per layer rises sharply, in some cases exceeding 60\%. The prefill stage can be even denser, in some cases approaching full activation. Even at small batch sizes, long prompts can activate many experts. Under these conditions, the expert working set can grow beyond the GPU budget, leading to frequent evictions and fetches. As shown in~\autoref{fig:waiting-vs-prompt}, we measure GPU stall latency as a function of input prompt length under ExpertFlow. As the number of input tokens increases, expert activation becomes denser and swap traffic can saturate PCIe bandwidth, introducing bubbles in an otherwise overlapped execution pipeline and leaving the GPU underutilized.

\begin{center}
\colorbox{gray!10}{%
\parbox{0.96\linewidth}{%
\textbf{Observation 1.}
MoE activation can become dense in prefill and under larger batch sizes, expanding the expert working set beyond what offloading and prefetching can reliably stage. When this happens, frequent evictions and fetches introduce GPU waiting time.}}
\end{center}

\begin{table}[t]
  \centering
  \caption{Expert activation ratio (\%) in decode stage.}
  \label{tab:expert_activation_ratio_decode}
  \vspace{-0.5em}
  \setlength{\tabcolsep}{4pt}
  \begin{tabularx}{\columnwidth}{@{}Xrrrrrr@{}}
    \toprule
    \textbf{Model} & \textbf{batch\_size=1} & \textbf{2} & \textbf{4} & \textbf{8} & \textbf{16} & \textbf{32} \\
    \midrule
    Qwen3-30B-A3B       & 6.3 & 11.6 & 20.1 & 31.9 & 46.5 & 62.0 \\
    Qwen3-Next-80B      & 1.9   & 3.7  & 6.7  & 14.9 & 25.4 & 39.2 \\
    DeepSeek-V2-Lite    & 9.0 & 14.9 & 23.2 & 35.9 & 51.7 & 67.6 \\
    \bottomrule
  \end{tabularx}
\end{table}

\begin{table}[t]
  \centering
  \caption{Expert activation ratio (\%) in prefill stage.}
  \label{tab:expert_activation_ratio_prefill}
  \vspace{-0.5em}
  \setlength{\tabcolsep}{4pt}
  \begin{tabularx}{\columnwidth}{@{}Xrrrrrr@{}}
    \toprule
    \textbf{Model} & \textbf{batch\_size=1} & \textbf{2} & \textbf{4} & \textbf{8} & \textbf{16} & \textbf{32} \\
    \midrule
    Qwen3-30B-A3B & 46.9 & 60.0 & 73.4 & 84.4 & 92.8 & 96.6 \\
    Qwen3-Next-80B   & 24.6 & 35.3 & 48.8 & 67.1 & 78.9 & 86.2 \\
    DeepSeek-V2-Lite   & 72.9 & 86.9 & 94.3 & 96.0 & 96.3 & 96.3 \\
    \bottomrule
  \end{tabularx}
\end{table}

\subsection{Post-Training Quantization for MoE}
\label{sec:background-ptq}

\begin{figure*}[h]
    \centering
    \begin{minipage}{0.32\linewidth}
        \centering
        \subfloat[Wikitext]{%
            \includegraphics[width=0.95\linewidth]{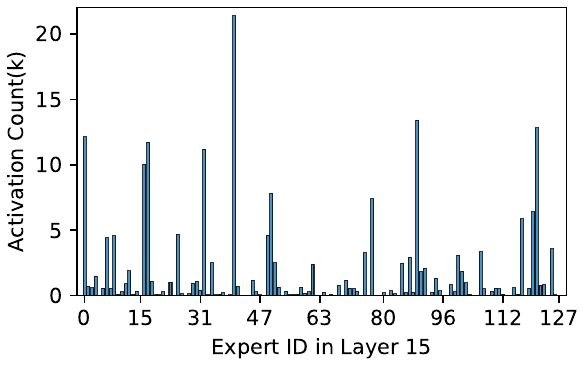}
        }
    \end{minipage}
    \begin{minipage}{0.32\linewidth}
        \centering
        \subfloat[GSM8K]{%
            \includegraphics[width=0.95\linewidth]{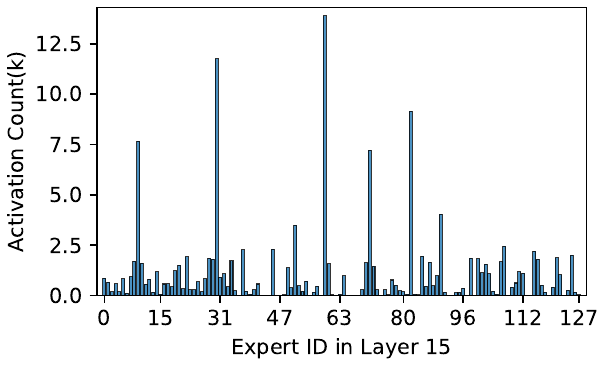}
        }
    \end{minipage}
    \begin{minipage}{0.32\linewidth}
        \centering
        \subfloat[HumanEval]{%
            \includegraphics[width=0.95\linewidth]{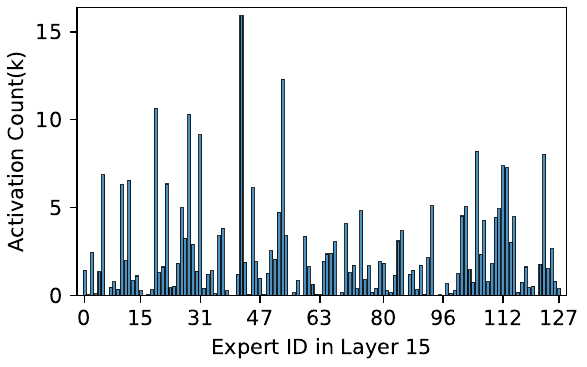}
        }
    \end{minipage}
    \caption{Expert activation counts under different workloads (Layer 15 of Qwen3-MoE-30B-A3B). Across three distinct workload types: WikiText (text), GSM8K (math), and HumanEval (code), the set of top-10 most frequently activated experts is entirely disjoint, highlighting pronounced routing shift across tasks.}
    \label{fig:expert-active}
    \vspace{-1.0em}
\end{figure*}

PTQ is a standard approach to reduce the memory footprint of LLMs without retraining by compressing weights to low bit-widths, often accompanied by calibration procedures that mitigate quantization error~\cite{lin2024awq,frantar2022gptq}. Recent work has extended PTQ beyond dense models to MoE architectures, where expert weights dominate the parameter budget and thus determine whether single-GPU deployment is feasible~\cite{kim2023mixturequantizedexpertsmoqe,frantar2023qmoepracticalsub1bitcompression,duanmu2025mxmoe,fu2025eaquantenhancingposttrainingquantization,zhang2025moqe,hu2025moequantenhancingquantizationmixtureofexperts}. Compared with dense-model PTQ, MoE PTQ must also contend with routing-induced imbalance: experts are activated unevenly, calibration data may be biased toward frequently selected experts, and heterogeneous expert execution can interact with kernel choices and memory traffic. Accordingly, several MoE-aware PTQ pipelines incorporate per-expert or per-block bit allocation, outlier handling, and in some cases system co-design for efficient execution of mixed-precision experts. Overall, this line of work substantially improves the quality--compression trade-off and makes it possible to run larger MoE models within constrained HBM.

Despite these advances, PTQ remains a lossy compression and often incurs a measurable accuracy gap to FP16, especially at aggressive bit-widths. For MoE serving, a more structural limitation is that most PTQ deployments either apply a uniform precision to all experts or fix a per-expert precision map offline and keep it unchanged at inference. This choice ignores a defining property of MoE execution: expert utilization is highly imbalanced over time. ~\autoref{fig:expert-active} shows a heavy-tailed distribution in cumulative activations, where a small hot set accounts for a large fraction of expert invocations while most experts remain cold. This does not contradict activation densification under larger batch sizes. Densification describes how many experts are activated concurrently within an iteration, whereas cumulative activations reflect long-horizon traffic concentration. Even when many experts are touched within the same prefill or decode step, their total usage over a serving window can still differ by orders of magnitude.

Hot-set identity is also workload dependent. ~\autoref{fig:expert-active} shows that for the same MoE model and layer, the top-10 most frequently activated experts under text, math, and code workloads can be entirely disjoint. Under such shifts, a static precision assignment may spend scarce memory budget on experts that contribute little traffic in the current workload, while over-compressing the experts that dominate execution. This misallocation is amplified under strict HBM budgets because each high-precision slot directly displaces memory that could have protected a currently hot expert.

\begin{center}
\colorbox{gray!10}{%
\parbox{0.96\linewidth}{%
\textbf{Observation 2 (Skew and shift in expert utilization).}
Over long horizons, MoE expert usage is heavy-tailed and workload dependent: a small hot set dominates cumulative invocations, and the hot-set identity can shift substantially across tasks. Static precision maps therefore misallocate scarce high-precision capacity under realistic serving mixes.}}
\end{center}

\subsection{Motivation for Online Precision Control}
\label{sec:background-motivation}

\begin{figure}[ht]
    \centering
    \begin{minipage}{0.46\linewidth}
        \centering
        \subfloat[Qwen3-30B-A3B]{%
            \includegraphics[width=0.99\linewidth]{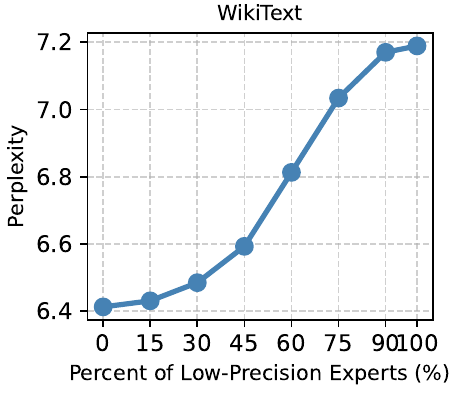}
        }
    \end{minipage}
    \begin{minipage}{0.46\linewidth}
        \centering
        \subfloat[Qwen3-80B-A3B]{%
            \includegraphics[width=0.99\linewidth]{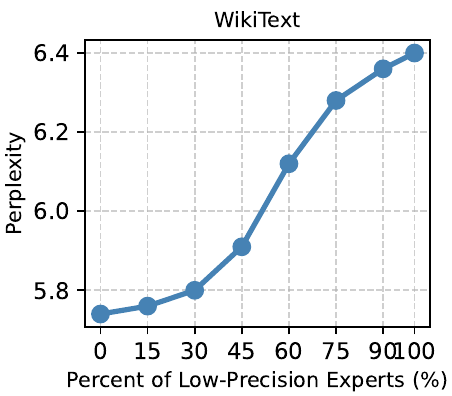}
        }
    \end{minipage}
    \caption{Perplexity Impact of Varying Low-Precision Expert Ratios Across Layers in \systemname\ (Qwen3-30B-A3B: FP16 vs. Int4, Qwen3-80B-A3B: Int4 vs. Int2)}
    \label{fig:ppl}
\end{figure}

The observations above suggest an opportunity to trade model quality for memory footprint under a strict GPU budget by adapting expert precision online. A key concern is whether such adaptation can be done without destabilizing quality. To probe this question, we quantify how language modeling quality changes as we increase the fraction of experts assigned to low precision. Using WikiText-2, we evaluate 128 prompts with 2048 input tokens and 256 output tokens, and progressively increase the number of demoted experts per layer. As Figure~\ref{fig:ppl} shows, when demotion is restricted to infrequently activated experts, perplexity increases smoothly as more experts are demoted. This behavior indicates that activation-aware precision assignment induces a predictable quality--compression curve,  and that protecting the frequently used experts captures a large portion of the quality benefit.

\begin{center}
\colorbox{gray!10}{%
\parbox{0.96\linewidth}{%
\textbf{Observation 3 (Smooth sensitivity to cold-expert demotion).}
When low precision is applied primarily to cold experts, increasing the number of quantified experts per layer yields a smooth and controllable increase in perplexity, suggesting a predictable quality--memory trade-off.}}
\end{center}

These results motivate \systemname, a runtime-aware precision allocation mechanism. \systemname\ uses routing traces as an online signal to continuously assign limited high-precision capacity to the experts that dominate current traffic, while compressing the rest under a fixed HBM budget.



\subsection{Challenges for Online Precision Allocation}
\label{sec:design-req}
Online precision control for MoE serving is challenging because it must adapt to routing shift while satisfying deployment constraints that are hard to relax:
\begin{itemize}[leftmargin=1.5em, itemsep=1pt, topsep=2pt]
    \item (i) \textbf{Hard budget invariance.} Precision changes directly alter the resident expert footprint. The system must never violate the HBM cap, otherwise OOM or forced evictions can cascade into instability.
    \item (ii) \textbf{Critical-path isolation.} Routing decisions occur on the token critical path, but precision transitions involve heavyweight device transfers. If transitions synchronize with the forward pass, they create stalls that dominate TTFT/TPOP.
    \item (iii) \textbf{Tail-latency robustness under non-stationarity.} Routing is workload-dependent, so naive reallocation can thrash when scores are close or fluctuate transiently, amplifying migration traffic and widening the P99 tail.
\end{itemize}
\section{Design: \systemname}
\label{sec:design}

To run MoE LLMs efficiently under a binding GPU memory budget, we designed \systemname\ to treat expert precision as an online, budget-constrained resource allocation problem. Each expert keeps two version: a low-precision version with bitwidth $b^{\text{lo}}$ and an high-precision version with bitwidth $b^{\text{hi}}$. At runtime, \systemname\ adapts which experts occupy $b^{\text{hi}}$ based on router-observed utilization, while enforcing three constraints: (C1) budget feasibility, (C2) non-blocking forward path, and (C3) stability under routing fluctuations.

\begin{figure*}[ht]
    \centering
    \includegraphics[width=0.93\linewidth]{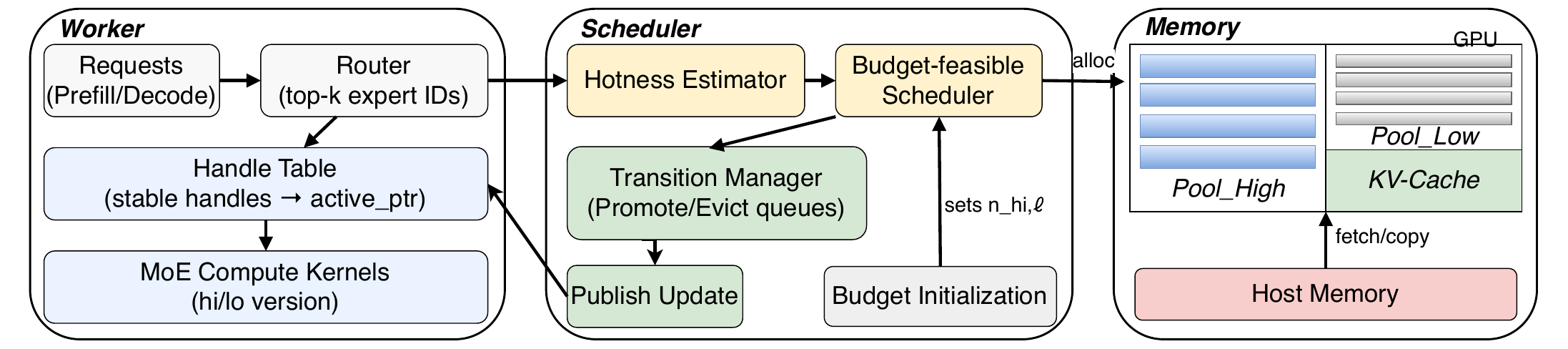}
    \caption{System architecture of \systemname.}
    \label{fig:architecture_overview}
    \vspace{-1em}
\end{figure*}

\subsection{Overview}
\label{sec:overview}

\autoref{fig:architecture_overview} summarizes the end-to-end control loop. The worker executes the MoE inference, while a scheduler observes router traces and updates expert precision residency under a strict GPU memory budget. The design separates the inference workflow from background transitions, and it exposes a stable interface so that residency changes do not perturb execution within a window.

On the worker side, each request invokes the router to obtain top-$k$ expert IDs, which are used to index a handle table. Each handle is stable in identity and resolves to an \texttt{active\_ptr} that points to a fully materialized expert version. 

Adaptation is driven by a scheduler that consumes router traces asynchronously. A hotness estimator maintains per-expert statistics over time. These statistics feed a budget-feasible scheduler that selects, for each layer $\ell$, a high-precision resident set subject to a layer capacity $n_{\mathrm{hi},\ell}$. The capacities are derived once by budget initialization, which accounts for fixed GPU allocations such as the KV cache and reserves the remaining device memory for expert weights. The resulting target sets drive a transition manager that maintains promote and evict queues and issues background work to materialize changes.

Memory is partitioned to make transitions predictable. On the GPU, expert weights are stored in separate pools for high- and low-precision versions, isolated from the KV cache region. When the transition manager promotes an expert, it allocates space from the high-precision pool and fetches the prepared weight version from host memory, then copies it to the device. After the copy completes, a publish step updates the handle table so that the new version becomes visible at the next iteration. Evictions reclaim the old version once the publish step is succeed. This organization aligns the architecture in Figure~\ref{fig:arch} with the four components described next: VER defines stable handles, memory management provides deterministic pools and budget gating, the transition pipeline performs non-blocking promote and evict operations with bounded interference, and the policy determines high-precision residency from router-derived hotness under the fixed per-layer capacities.

\subsection{Versioned Expert Residency (VER)}
\label{sec:ver}

To manage experts with different precisions on the GPU and enable expert precision switching when the workload or activation distribution changes, we designed Versioned Expert Residency (VER). VER provides a mechanism for controlling expert residency through versioned precision. For each MoE layer $\ell$ and expert $e$, VER maintains multiple weight versions that implement the same operator in different numeric formats. In the simplest configuration, each expert has a high-precision version and a low-precision version. The high-precision version targets accuracy-critical execution, while the low-precision version provides a memory-efficient fallback.

\paragraph{Expert entries and stable handles.}
VER represents each expert as an \emph{expert entry} that owns metadata for all supported versions. Each entry exports a \emph{stable handle} that is passed to the MoE kernel at execution time. The handle is immutable in identity but contains a pointer to the currently active version on the GPU. The compute path resolves the handle to obtain the active pointer and the associated quantization parameters, then invokes the corresponding kernel. This indirection allows VER to change residency and precision by updating the active pointer while keeping the handle location stable. As a result, version changes do not require updating kernel arguments beyond the handle itself.

\paragraph{Residency states.}
Each expert entry tracks the residency of its versions on the GPU. We consider four states. In \textsc{Resident-Hi}, the high-precision version is present on the GPU and the handle points to it. In \textsc{Resident-Lo}, only the low-precision version is present and the handle points to it. In \textsc{Promoting}, the system is transferring the high-precision version to the GPU; during this state, the handle continues to point to a previously valid version, typically the low-precision one. In \textsc{Demoting}, the system is transferring the low-precision version to the GPU to replace the current high-precision version. In \textsc{Evicting}, the system reclaims the old version storage after switching the handle to a remaining valid version. These states enforce a single invariant: the handle must always resolve to a complete and usable weight version. This invariant is sufficient to keep the forward pass non-blocking even when promotions/demotions are in flight.

\paragraph{Non-blocking switching semantics.}
VER decouples version transitions from the token critical path. When the runtime decides that an expert should be promoted or demoted, it initiates a background transfer to allocate GPU space and copy the high-precision weights. The forward pass does not wait for this transfer. Instead, it continues to use the current active version via the stable handle. After the transfer completes, VER atomically publishes the new version by updating the handle's active pointer. Similarly, eviction proceeds by first redirecting the handle to a still-resident version and then reclaiming the freed storage in the background. This publish-then-switch discipline ensures that no kernel observes a partially populated version.

VER isolates residency changes behind stable handles and enforces that the forward pass always has a valid version to execute. This separation allows the scheduling policy to focus on deciding which experts should occupy high-precision capacity under a memory budget. In the remainder of the design, we describe how VER is combined with deterministic memory management and an online scheduler to bound transition overheads and control interference on shared memory.

\subsection{GPU Memory Management}
\label{sec:memmgmt}

Dynamic expert residency stresses the GPU allocator in two ways. First, promotions and evictions introduce frequent allocations for large weight buffers, which can fragment the address space and increase allocation latency variance. Second, weight transitions often require temporary staging buffers whose peak demand depends on the concurrency of in-flight promotions. If these allocations share the same allocator path as the compute stack, transient pressure can propagate into the token critical path through allocator jitter and incidental synchronization. We therefore manage memory with explicit partitions and fixed-granularity allocation, and we gate every transition by a global budget tracker.

\paragraph{Partitioned pools.}
We divide the GPU memory region assigned to expert weights into disjoint pools with non-overlapping roles. The high-precision pool $\texttt{pool\_hi}$ and low-precision pool $\texttt{pool\_lo}$ store resident high-/low-precision expert versions. These pools are responsible for the largest allocations and are the primary source of fragmentation if left unmanaged.

\paragraph{Fixed-granularity allocation.}
$\texttt{pool\_hi}$ and $\texttt{pool\_lo}$ allocate memory in fixed-size blocks and serve requests by composing one or more blocks. Block size is chosen to balance internal fragmentation and allocation overhead; in our implementation we align blocks to large granularities comparable to expert size so that allocation and reclamation remain predictable. Each pool maintains a constant-time free list, so allocation and free are simple pointer operations without invoking the general-purpose runtime allocator. This design eliminates allocator contention.

\paragraph{Budget model and OOM safety.}
We enforce a global memory budget through a \texttt{BudgetTracker} that exposes explicit reservation and release operations. Let $M_{\text{total}}$ denote the usable GPU memory for the process. We reserve a fixed portion $M_{\text{fixed}}$ for non-expert components, including the KV cache, non-expert parameters, activations, and framework runtime overhead. The remaining capacity is split into an upper bound for high-precision expert residency, $M_{\text{exp,hi}}^{\text{cap}}$, and low-precision expert residency $M_{\text{tmp}}^{\text{cap}}$. 

Every promotion/demotion request must pass an admission check before entering the transition pipeline. Given a candidate expert whose high-precision version requires $b_{\text{hi}}$ bytes in \texttt{pool\_hi}, the runtime calls \texttt{try\_reserve}$(b_{\text{hi}})$ on the \texttt{BudgetTracker}. A successful reservation guarantees that subsequent allocation from \texttt{pool\_hi} will not trigger out-of-memory failures, and the promotion proceeds asynchronously. If the reservation fails, the runtime does not initiate the promotion. Instead, it defers the promotion or schedules evictions to reclaim capacity, while the forward path continues to execute using the pinned low-precision version in the current window. 

\subsection{Non-blocking Transition Pipeline}
\label{sec:pipeline}

VER decouples version visibility from data movement using window-level pinning, but the system still needs a concrete pipeline that prepares promotions and performs evictions without stalling the forward pass. The transition pipeline serves two purposes. It keeps the compute stream independent from weight transfers by running transitions on a dedicated migration stream, and it enforces backpressure so that background activity does not amplify tail latency through bandwidth contention.

\paragraph{Asynchronous queues and backpressure.}
The runtime maintains two logical queues, an update queue for promotion and demotion and an eviction queue, each containing expert identifiers $(\ell,e)$ that are candidates for residency changes. A background worker consumes these queues and issues asynchronous work. The worker prioritizes evictions when the memory budget is tight, since reclaiming high-precision buffers increases the feasible set of subsequent updates. Before launching an update, the worker ensures that the temporary pool has sufficient capacity for staging buffers and that the global budget tracker admits the high-precision allocation. Promotions and evictions are executed on a dedicated CUDA stream, denoted $\texttt{stream}_{\text{mig}}$, which is disjoint from the compute stream used by attention and expert kernels. This separation avoids implicit synchronization on the critical path and makes transition overheads observable and controllable.

Backpressure is implemented at admission: a promotion is enqueued only if it passes the budget reservation check described in Section~\ref{sec:memmgmt}. When $\texttt{pool\_hi}$ and $\texttt{pool\_lo}$ are full, additional promotions remain queued. The forward pass continues to execute using the currently pinned mapping.

\begin{figure}[t]
    \centering
    \includegraphics[width=0.99\columnwidth]{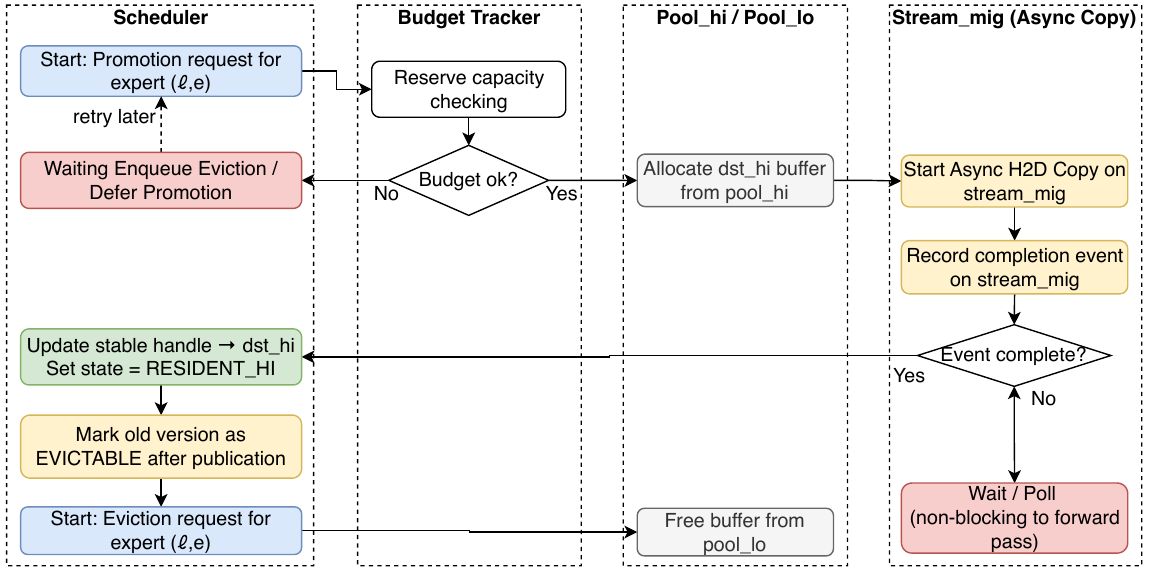}
    \caption{Procedure of Update/Eviction.}
    \label{fig:procedure}
    \vspace{-1em}
\end{figure}

\paragraph{Update/Evict procedure.}
Taking the promotion process as an example, it materializes the high-precision version of an expert in GPU memory and makes it visible only at a safe publication point. For a target expert $(\ell,e)$, the worker first reserves the required capacity from the global budget tracker. It then allocates the destination buffer from $\texttt{pool\_hi}$. Next, it issues an asynchronous copy on $\texttt{stream}_{\text{mig}}$ to populate the destination with the pre-packed high-precision weights. The source resides in host memory; the design avoids on-the-fly repacking during promotion to prevent unpredictable temporary allocations and extra bandwidth pressure. After the copy is issued, the worker records a completion event on $\texttt{stream}_{\text{mig}}$. The expert entry remains in the promoting state until this event is complete. Publication updates the stable handle to point to the new high-precision buffer. Publication is performed only after the completion event, ensuring that the forward pass never observes a partially populated version. 

As for the eviction procedure, it reclaims high-precision / low-precision buffers after they are no longer required by the pinned mapping. The worker marks the corresponding old version buffer as evictable after publication completes.

\subsection{Online Scheduling Policy}
\label{sec:policy}

VER and the transition pipeline define how expert versions are materialized and published under window-level pinning. The remaining question is which experts should occupy the limited high-precision capacity at each layer. We use a lightweight policy driven by router traces. The policy is intentionally simple so that its overhead is negligible and its behavior is easy to reason about under workload shifts.

\paragraph{Hotness Estimation}
For each MoE layer $\ell$ and expert $e$, the runtime maintains a counter $c_{\ell,e}$ that accumulates the number of times $(\ell,e)$ is selected by the router during the current update interval. Updates are triggered periodically with a fixed interval $T_u$ measured in time. A time-based interval is stable under varying batch composition and prompt lengths because it does not depend on the number of tokens processed within the interval.

At the end of each update interval, the runtime updates a smoothed hotness score $S_{\ell,e}$ using an exponential moving average,
\[
S_{\ell,e} \leftarrow \alpha \cdot S_{\ell,e} + (1-\alpha)\cdot c_{\ell,e},
\]
where $\alpha \in [0,1)$ controls the tradeoff between responsiveness and stability. The counters $c_{\ell,e}$ are then reset for the next interval. This construction uses router outputs only and does not assume access to labels or online quality signals.

\paragraph{Budget-feasible Selection}
High-precision capacity is fixed by the memory budget. For each layer $\ell$, let $n_{\mathrm{hi},\ell}$ denote the maximum number of experts whose high-precision versions can be resident concurrently. Given the hotness scores, the policy selects the high-precision set as the top-$n_{\mathrm{hi},\ell}$ experts,
\[
H_\ell \leftarrow \mathrm{TopN}\!\left(\{S_{\ell,e}\}_e,\; n_{\mathrm{hi},\ell}\right).
\]
This operation is local to each layer and is budget-feasible by construction. It yields an explicit target set for the next window and avoids global optimization that would require expensive coordination across layers.

The resulting set difference determines which experts change residency. Let $H_\ell^{\mathrm{cur}}$ denote the high-precision set currently pinned in the active window. The candidates for promotion are $H_\ell \setminus H_\ell^{\mathrm{cur}}$, and the candidates for demotion are $H_\ell^{\mathrm{cur}} \setminus H_\ell$. These candidates are passed to the transition pipeline, which admits promotions only when the budget tracker can reserve the required memory.

\paragraph{Hysteresis for Stability}
A naive top-$N$ rule can cause unnecessary churn when several experts have similar hotness scores. Such churn increases the number of promotions and amplifies background bandwidth consumption without commensurate quality gains. We therefore apply hysteresis when updating $H_\ell$.

Concretely, an expert is promoted into the high-precision set only if its hotness exceeds that of the lowest-ranked expert in the current high-precision set by a margin. Symmetrically, an expert is evicted only if its hotness falls below the highest-ranked expert outside the current set by the same margin. This margin can be expressed as an additive threshold on the score or as a rank slack that requires an expert to enter a slightly wider candidate set before it becomes eligible. Hysteresis does not change the budget constraint. It reduces oscillations and makes the transition rate more predictable under transient routing fluctuations.

The policy operates at the update cadence $T_u$ and produces a target high-precision set for each layer. Window-level pinning then determines when these targets become visible to the compute path, and the transition pipeline enforces admission control and bounded interference when materializing new high-precision residents.

\section{Implementation}
\label{sec:impl}

We implement the design on top of the HuggingFace Transformers stack as a self-contained Python extension of roughly 3K lines that modifies the MoE execution path while preserving the original model interfaces. The implementation instruments the router to expose per-token top-$k$ expert IDs and aggregates these traces on the host for hotness estimation and windowed scheduling. VER is realized by a persistent handle table that maps each expert to a version metadata. We implement deterministic memory management as fixed-granularity device pools for high-precision weights and temporary staging buffers, together with a budget tracker that reserves capacity before admitting promotions. The transition manager runs in a background thread and issues asynchronous host-to-device copies on a dedicated CUDA stream, using CUDA events to detect copy completion before publishing updates. Expert weights are prepared offline into high- and low-precision versions in kernel-ready layouts and stored in pinned host memory; at runtime the system promotes and evicts experts by updating residency and the handle table while the forward pass continues to execute on the compute stream with the currently pinned mapping.
\section{Evaluation}


We evaluate \systemname\ under a single-GPU device-memory budget, focusing on the tradeoffs among quality, latency, and throughput. The evaluation addresses two questions.
\begin{itemize}[leftmargin=1.5em, itemsep=1pt, topsep=2pt]
    \item {Q1: Quality.} At the same device-memory footprint, how much quality does online, hotness-driven precision allocation recover relative to static compression? (~\autoref{sec:eval-quality})
    \item {Q2: Performance.} When device memory is the binding constraint, what latency and throughput does \systemname\ achieve? (~\autoref{sec:eval-performance})
\end{itemize}

\subsection{Experimental Setup}
\label{sec:eval-setup}

\begin{table}[t]
  \centering
  \caption{Configuration of evaluated MoE models.}
  \label{tab:moe_config}
  \vspace{-0.5em}
  \setlength{\tabcolsep}{4pt}
  \resizebox{\linewidth}{!}{
  \begin{tabular}{lccc}
    \toprule
    & \textbf{Qwen3-30B} & \textbf{Qwen3-80B (Int4)} & \textbf{Phi-MoE} \\
    \midrule
    Total Parameters & 30.5B & 80B & 42B \\
    Activated Parameters/Token & 3.3B & 3B & 6.6B \\
    Total Weight Size & 57GB & 41GB & 78GB \\
    Experts Weight Size & 54GB(95\%) & 37GB(93\%) & 75GB(96\%) \\
    Layer Number & 48 & 48 & 32 \\
    Expert Number/Layer & 128 & 512 & 16 \\
    Shared Experts/Layer & 0 & 1 & 0 \\
    Top-$K$ & 8 & 10 & 2 \\
    \bottomrule
  \end{tabular}}
\end{table}

\noindent\textbf{Models.}
We evaluate \systemname\ on Qwen3-30B-A3B-Instruct, Qwen3-80B-A3B-Instruct~\cite{qwen3technicalreport} and Phi-3.5-MoE-instruct~\cite{abdin2024phi3}. The detail of these model are shown in~\autoref{tab:moe_config}.

\noindent\textbf{Benchmarks and quality metrics.}
We report task performance on WikiText-2~\cite{merity2016pointer} (perplexity), MMLU-Pro~\cite{wang2024mmlu}, GPQA~\cite{rein2024gpqa}, AIME25~\cite{balunovic_srimatharena_2025}, GSM8K~\cite{cobbe2021gsm8k}, and HumanEval~\cite{chen2021evaluating}.

\noindent\textbf{Serving stack and hardware.}
We integrate \systemname\ into a PyTorch and Transformers based MoE inference stack~\cite{paszke2019pytorch,wolf2020transformers}. And We run all experiments on a single RTX A6000 (48\,GB), and we uses two precision tiers in our experiments: Hot experts are assigned a higher-precision version(usually FP16, while INT4 for Qwen3-80B), and cold experts remain in a lower-precision version(INT4 or INT2).

\noindent\textbf{Latency and throughput under resource limits.}
We evaluate serving performance with three complementary measurements. First, we sweep batch size and report TTFT and TPOP for both average and P99. Second, we vary the number of tokens and measure how latency scales with prompt length and generation length, again reporting average and P99. Third, we report end-to-end throughput for both prefill and decode. All measurements are performed under the same memory budget, so the reported tail behavior reflects the interaction between background transitions and the compute path.

\subsection{Quality of \systemname}
\label{sec:eval-quality}

\begin{table*}[htbp]
\centering
\caption{Comparison of accuracy across models/methods.}
\label{tab:accuracy}
\vspace{-0.5em}
\begin{tabular}{llrrrrrrr}
\toprule
\textbf{MODEL} & \textbf{METHOD} & \textbf{MMLU-Pro} & \textbf{GPQA} & \textbf{AIME25} & \textbf{GSM8K} & \textbf{Human Eval} & \textbf{AVG.} \\
\midrule
Qwen3-MoE-30B & FP16    & 73.37 & 54.55 & 23.33 & 90.45 & 84.76 & 65.29 \\
              & Int4    & 72.84 & 53.54 & 20.00 & 89.39 & 84.15 & 63.98 \\
              & DynaExq & 73.09 & 54.14 & 20.00 & 89.91 & 84.76 & \textbf{64.38} \\
\midrule
Qwen3-MoE-80B & Int4    & 75.92 & 72.22 & 70.00 & 87.04 & 85.37 & 78.11 \\
              & Int2    & 72.75 & 66.67 & 63.33 & 80.97 & 81.71 & 73.09 \\
              & DynaExq & 75.48 & 71.21 & 66.67 & 86.73 & 84.76 & \textbf{77.57} \\
\midrule
Phi-3.5-MoE & FP16  & 54.23 & 36.75 & 40.00 & 88.66 & 70.64 & 58.06 \\
            & Int4  & 53.40 & 34.52 & 36.67 & 87.54 & 69.41 & 56.31 \\
            & DynaExq & 53.91 & 35.94 & 36.67 & 88.25 & 70.22 & \textbf{57.00} \\

\midrule
\bottomrule
\end{tabular}
\end{table*}

\begin{figure*}[t]
    \centering
    \begin{minipage}{0.3\linewidth}
        \centering
        \subfloat[Qwen3-30B Avg TTFT]{%
            \includegraphics[width=0.99\linewidth]{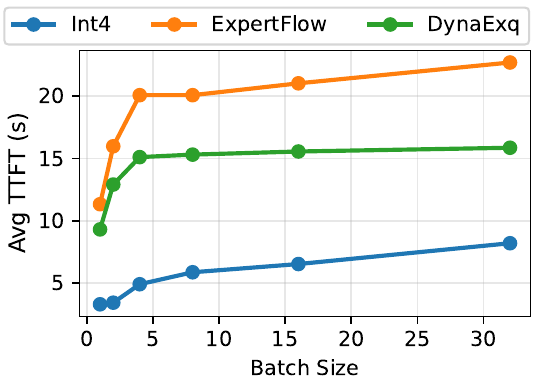}
        }
    \end{minipage}
    \begin{minipage}{0.3\linewidth}
        \centering
        \subfloat[Qwen3-30B P99 TTFT]{%
            \includegraphics[width=0.99\linewidth]{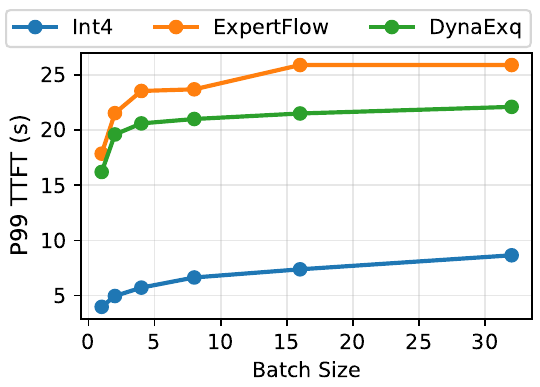}
        }
    \end{minipage}
    \begin{minipage}{0.3\linewidth}
        \centering
        \subfloat[Qwen3-80B Avg TTFT]{%
            \includegraphics[width=0.99\linewidth]{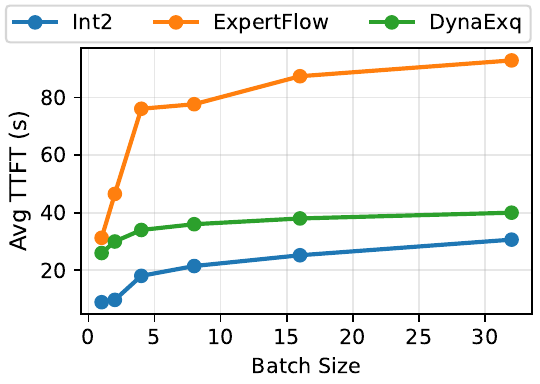}
        }
    \end{minipage}
    \begin{minipage}{0.3\linewidth}
        \centering
        \subfloat[Qwen3-80B P99 TTFT]{%
            \includegraphics[width=0.99\linewidth]{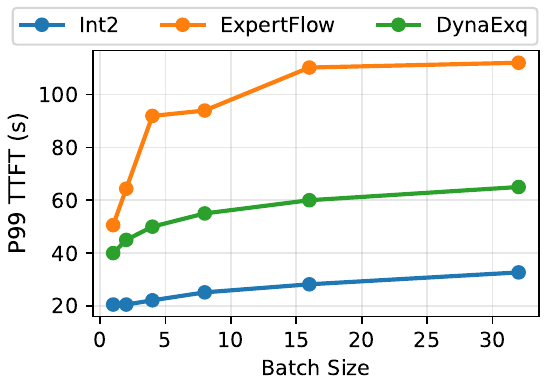}
        }
    \end{minipage}
    \begin{minipage}{0.3\linewidth}
        \centering
        \subfloat[Phi-3.5-MoE Avg TTFT]{%
            \includegraphics[width=0.99\linewidth]{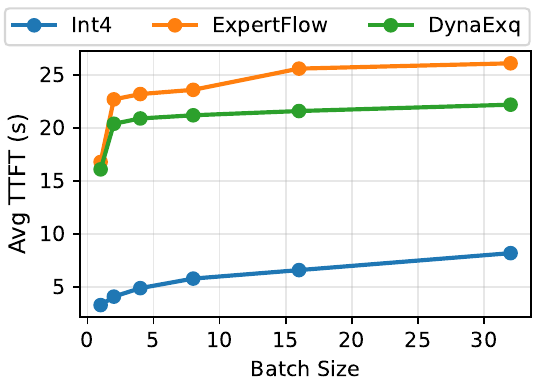}
        }
    \end{minipage}
    \begin{minipage}{0.3\linewidth}
        \centering
        \subfloat[Phi-3.5-MoE P99 TTFT]{%
            \includegraphics[width=0.99\linewidth]{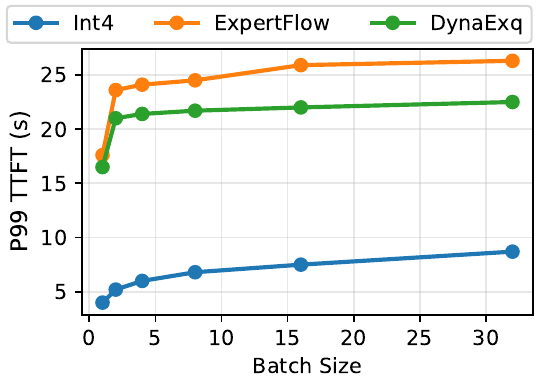}
        }
    \end{minipage}
    \caption{The comparison of TTFT across different batch size.}
    \label{fig:ttft}
    \vspace{-1em}
\end{figure*}

Table~\ref{tab:accuracy} compares \systemname\ with FP16 and static low-bit quantization. On Qwen3-MoE-30B, static Int4 reduces the average score from 65.29 (FP16) to 63.98. With the same single GPU capability, \systemname\ raises the average to 64.38. The gain comes from improvements on GPQA (54.14 vs.\ 53.54), GSM8K (89.91 vs.\ 89.39), and HumanEval (84.76 vs.\ 84.15). These results indicate that reallocating high precision to persistently used experts can recover part of the quality loss without increasing the model footprint.

The benefit is larger when the static baseline is forced to a more aggressive bitwidth to fit the budget. On Qwen3-MoE-80B, Int2 degrades the average to 73.09, whereas \systemname\ improves it to 77.57 under the same budget. The recovery is consistent across benchmarks. Compared to the Int4 baseline, \systemname\ remains close in aggregate (77.57 vs.\ 78.11), suggesting that dynamic allocation can approach a higher-bit configuration when capacity constraints would otherwise require uniform low-bit compression.

We observe a similar pattern on Phi-3.5-MoE. Static Int4 lowers the average from 58.06 to 56.31, while \systemname\ increases it to 57.00. The improvements are reflected on GPQA (35.94 vs.\ 34.52), GSM8K (88.25 vs.\ 87.54), and HumanEval (70.22 vs.\ 69.41). Overall, these results support the premise that router-driven precision allocation can preserve accuracy under a fixed GPU budget by concentrating higher precision on the experts that account for a larger share of execution.

\subsection{Performance of \systemname.}
\label{sec:eval-performance}

We evaluate performance on a single A6000 GPU under the same device-memory budget across methods. We compare \systemname\ against a static quantization baseline (Int4 for Qwen3-30B and Phi-3.5-MoE; Int2 for Qwen3-80B) and a offloading and prefetching system: ExpertFlow±\cite{shen2025expertflow}. We report prefill time-to-first-token (TTFT), decode time-per-output-token (TPOP), end-to-end request latency, and throughput. To stress the regimes where MoE execution becomes less sparse, we sweep batch size and, correspondingly, increase the number of tokens processed per iteration. We report both average and P99 to capture tail behavior under contention.

\paragraph{\textbf{TTFT}}
\autoref{fig:ttft} shows TTFT as batch size increases. The static quantization baseline provides the lowest TTFT, since it avoids weight movement on the critical path. ExpertFlow incurs a sharp increase in both average and P99 TTFT as batch grows. This trend is consistent with prefill activating a larger fraction of experts, which increases transfers and reduces overlap opportunities, turning migration into visible waiting time. \systemname\ tracks the static baseline more closely and remains substantially below ExpertFlow across batch sizes. The gap is larger at higher batch sizes, where prefill becomes effectively dense and offloading policies face sustained pressure from repeated promotions.

\begin{figure*}[t]
    \centering
    \begin{minipage}{0.3\linewidth}
        \centering
        \subfloat[Qwen3-30B Avg TPOP]{%
            \includegraphics[width=0.99\linewidth]{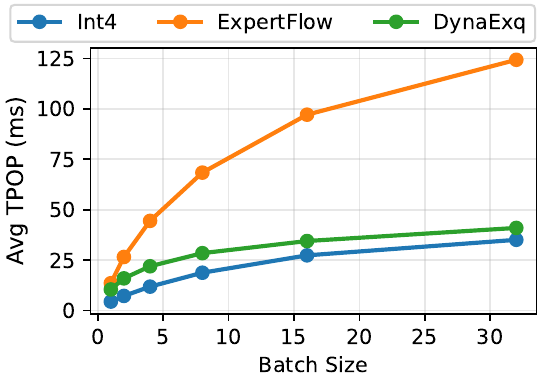}
        }
    \end{minipage}
    \begin{minipage}{0.3\linewidth}
        \centering
        \subfloat[Qwen3-30B P99 TPOP]{%
            \includegraphics[width=0.99\linewidth]{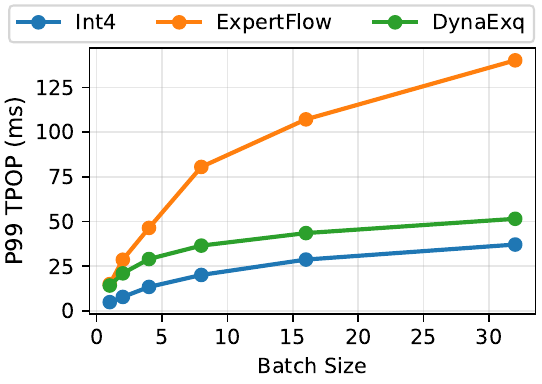}
        }
    \end{minipage}
    \begin{minipage}{0.3\linewidth}
        \centering
        \subfloat[Qwen3-80B Avg TPOP]{%
            \includegraphics[width=0.99\linewidth]{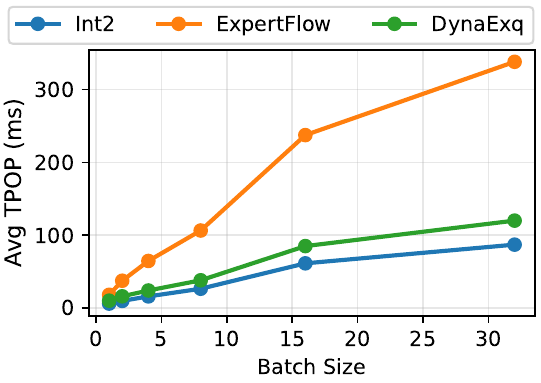}
        }
    \end{minipage}
    \begin{minipage}{0.3\linewidth}
        \centering
        \subfloat[Qwen3-80B P99 TPOP]{%
            \includegraphics[width=0.99\linewidth]{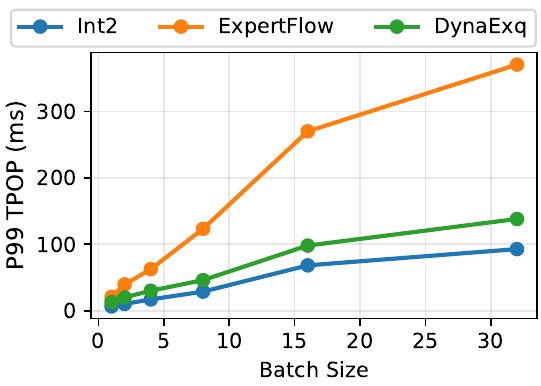}
        }
    \end{minipage}
    \begin{minipage}{0.3\linewidth}
        \centering
        \subfloat[Phi-3.5-MoE Avg TPOP]{%
            \includegraphics[width=0.99\linewidth]{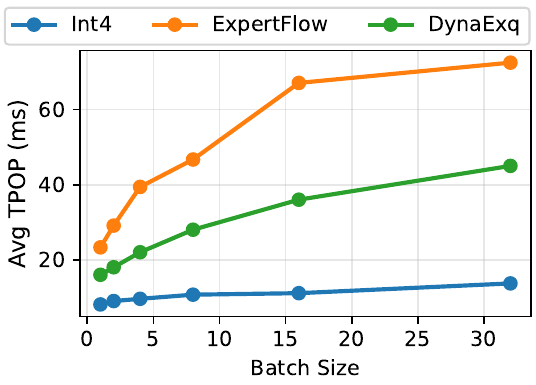}
        }
    \end{minipage}
    \begin{minipage}{0.3\linewidth}
        \centering
        \subfloat[Phi-3.5-MoE P99 TPOP]{%
            \includegraphics[width=0.99\linewidth]{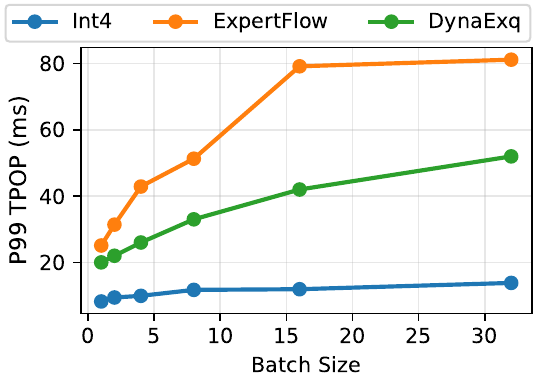}
        }
    \end{minipage}
    \caption{The comparison of TPOP across different batch size.}
    \label{fig:tpop}
    \vspace{-1em}
\end{figure*}

\paragraph{\textbf{TPOP}}
\autoref{fig:tpop} reports TPOP during decode. Decode is less sensitive to prompt-length effects, but it is still affected by background migration when the expert working set changes across concurrent requests. ExpertFlow again shows higher TPOP and a widening tail as batch increases, indicating that transfer interference is not confined to prefill. \systemname\ reduces this interference by separating the compute and migration streams and bounding migration rate, so its TPOP remains close to static quantization, with a smaller separation between average and P99.

\paragraph{\textbf{End-to-End Latency}}
\autoref{fig:end2end} summarizes end-to-end latency. The ordering mirrors TTFT and TPOP: static quantization is lowest, ExpertFlow is highest, and \systemname\ lies in between while remaining closer to the static baseline. As the per-iteration token volume increases with batch size, ExpertFlow experiences compounding delays from repeated transfers and synchronization, which are reflected in both average and P99 latency. \systemname\ avoids blocking on transitions and limits background interference, so the end-to-end latency curve grows more gradually. The same pattern carries to throughput: by keeping prefill and decode from stalling on expert movement, \systemname\ sustains higher effective throughput than ExpertFlow at larger token volumes, while staying near the static baseline under the same memory budget.

\begin{figure*}[t]
    \centering
    \begin{minipage}{0.3\linewidth}
        \centering
        \subfloat[Qwen3-30B Avg Latency]{%
            \includegraphics[width=0.99\linewidth]{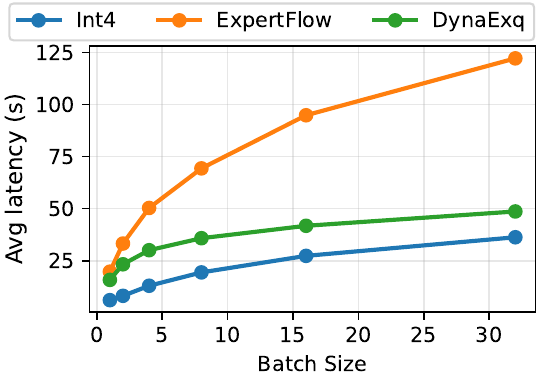}
        }
    \end{minipage}
    \begin{minipage}{0.3\linewidth}
        \centering
        \subfloat[Qwen3-30B P99 Latency]{%
            \includegraphics[width=0.99\linewidth]{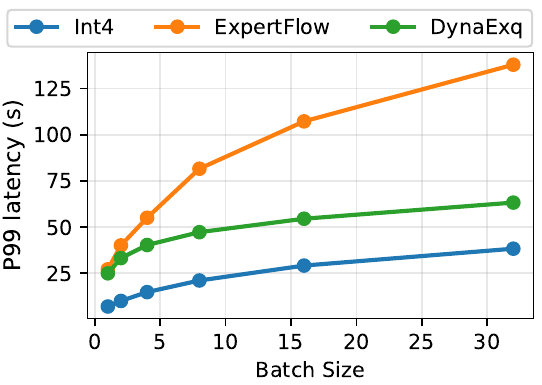}
        }
    \end{minipage}
    \begin{minipage}{0.3\linewidth}
        \centering
        \subfloat[Qwen3-80B Avg Latency]{%
            \includegraphics[width=0.99\linewidth]{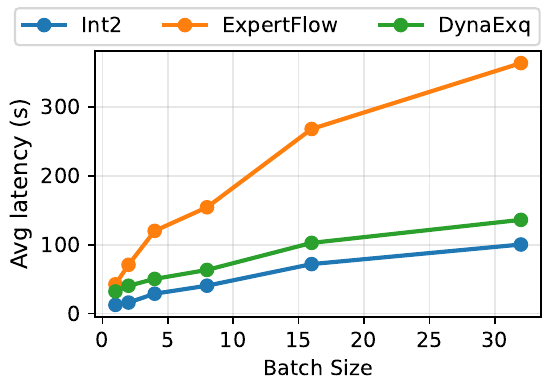}
        }
    \end{minipage}
    \begin{minipage}{0.3\linewidth}
        \centering
        \subfloat[Qwen3-80B P99 Latency]{%
            \includegraphics[width=0.99\linewidth]{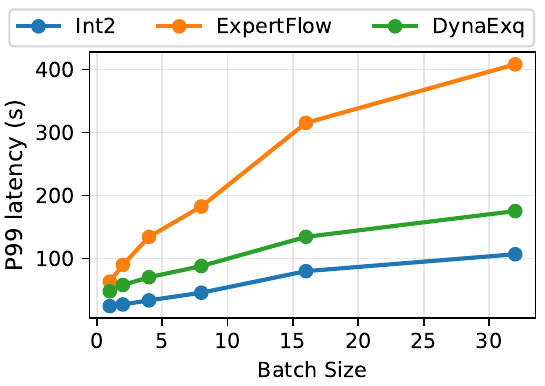}
        }
    \end{minipage}
    \begin{minipage}{0.3\linewidth}
        \centering
        \subfloat[Phi-3.5-MoE Avg Latency]{%
            \includegraphics[width=0.99\linewidth]{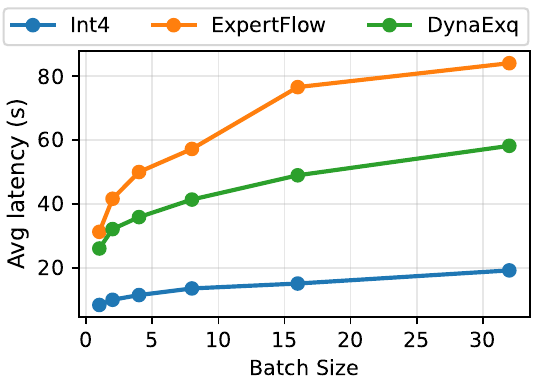}
        }
    \end{minipage}
    \begin{minipage}{0.3\linewidth}
        \centering
        \subfloat[Phi-3.5-MoE P99 Latency]{%
            \includegraphics[width=0.99\linewidth]{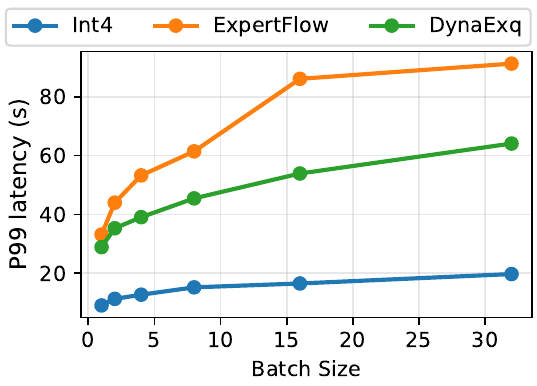}
        }
    \end{minipage}
    \caption{The comparison of End-to-End Latency across different batch size.}
    \label{fig:end2end}
    \vspace{-1em}
\end{figure*}

\paragraph{\textbf{Throughput under increasing batch size}}
\autoref{fig:throughput} reports end-to-end throughput in tokens/s as we increase batch size on a single A6000 under the same device-memory budget. Across all three models, \systemname\ consistently outperforms ExpertFlow, with throughput improvements ranging from $1.42\times$ to $2.73\times$. The gap widens as batch size grows. At higher batch sizes, prefill activates a larger fraction of experts within each iteration, which increases the amount of expert movement and reduces overlap opportunities for offloading and prefetching. This effect limits ExpertFlow's scaling and leads to early saturation. In contrast, \systemname\ keeps the high-precision working set resident within the fixed budget and bounds background transition interference, so throughput increases more steadily with batch size. The trend holds for Qwen3-30B, Qwen3-80B, and Phi-3.5-MoE, indicating that the benefit is not tied to a specific expert pool size but to reducing transfer-induced stalls when the per-iteration token volume increases.

\begin{figure*}[t]
    \centering
    \begin{minipage}{0.3\linewidth}
        \centering
        \subfloat[Qwen3-30B Throughput]{%
            \includegraphics[width=0.99\linewidth]{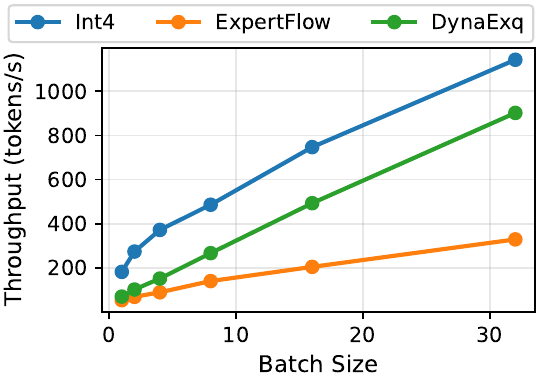}
        }
    \end{minipage}
    \begin{minipage}{0.3\linewidth}
        \centering
        \subfloat[Qwen3-80B Throughput]{%
            \includegraphics[width=0.99\linewidth]{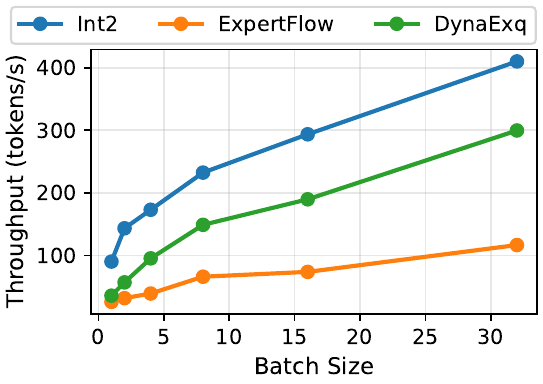}
        }
    \end{minipage}
    \begin{minipage}{0.3\linewidth}
        \centering
        \subfloat[Phi-3.5-MoE Throughput]{%
            \includegraphics[width=0.99\linewidth]{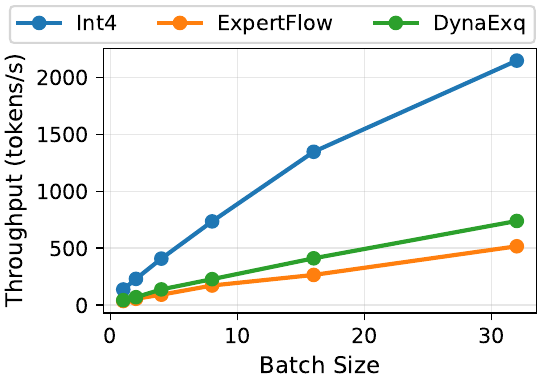}
        }
    \end{minipage}
    \caption{The comparison of End-to-End Throughput (tokens/s) across different batch size.}
    \label{fig:end2end}
    \vspace{-1em}
\end{figure*}

\begin{figure*}[t]
    \centering
    \begin{minipage}{0.3\linewidth}
        \centering
        \subfloat[Qwen-30B Avg Latency]{%
            \includegraphics[width=0.99\linewidth]{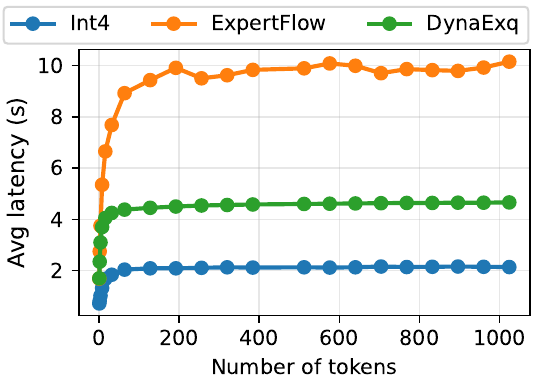}
        }
    \end{minipage}
    \begin{minipage}{0.3\linewidth}
        \centering
        \subfloat[Qwen-80B Avg Latency]{%
            \includegraphics[width=0.99\linewidth]{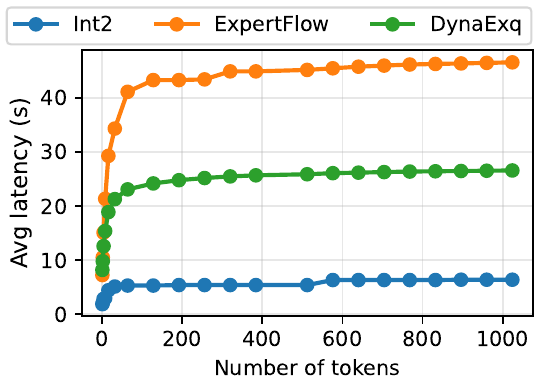}
        }
    \end{minipage}
    \begin{minipage}{0.3\linewidth}
        \centering
        \subfloat[Phi-3.5-MoE Avg Latency]{%
            \includegraphics[width=0.99\linewidth]{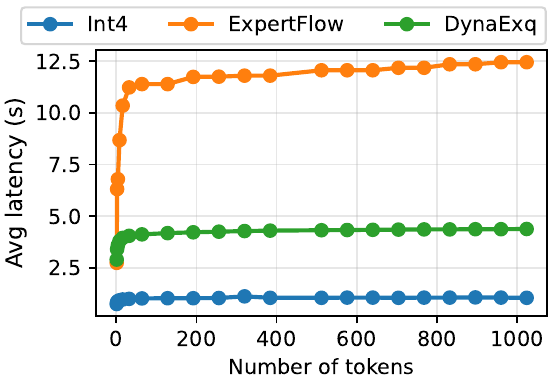}
        }
    \end{minipage}
    \begin{minipage}{0.3\linewidth}
        \centering
        \subfloat[Qwen-30B P99 Latency]{%
            \includegraphics[width=0.99\linewidth]{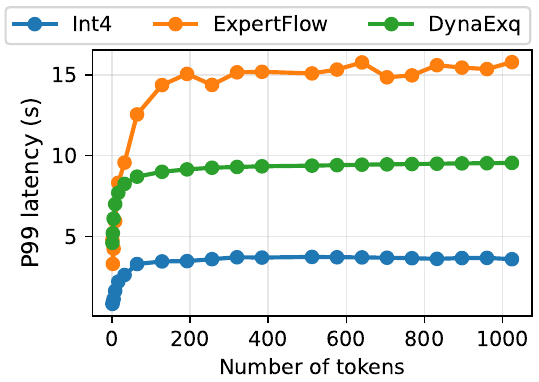}
        }
    \end{minipage}
    \begin{minipage}{0.3\linewidth}
        \centering
        \subfloat[Qwen-80B P99 Latency]{%
            \includegraphics[width=0.99\linewidth]{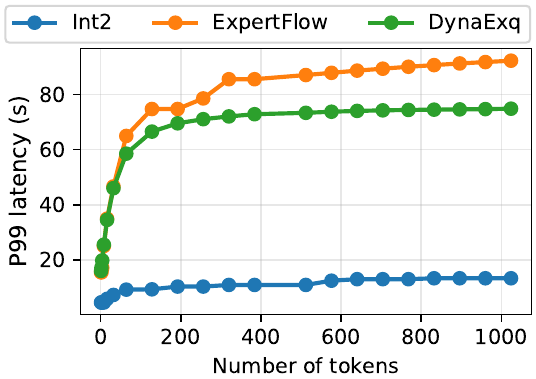}
        }
    \end{minipage}
    \begin{minipage}{0.3\linewidth}
        \centering
        \subfloat[Phi-3.5-MoE P99 Latency]{%
            \includegraphics[width=0.99\linewidth]{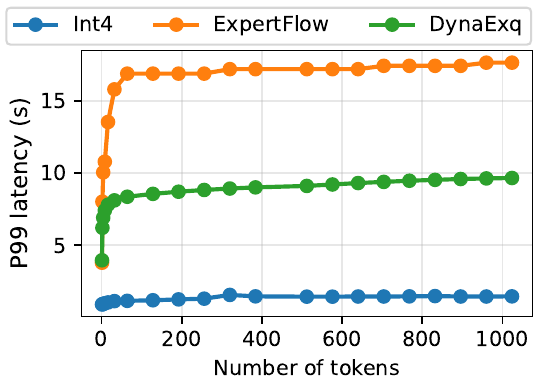}
        }
    \end{minipage}
    \caption{The comparison of TTFT latency across different prompt length.}
    \label{fig:ttft_tokens}
\end{figure*}

\paragraph{\textbf{Latency scaling with prompt length}}
\autoref{fig:ttft_tokens} further sweeps the number of tokens processed per request and reports the resulting TTFT latency (average and P99). Across models, latency increases quickly when the prompt grows from very short inputs to a few hundred tokens, and then approaches a plateau as the per-request prefill work dominates. The static quantization baselines remain the lowest and change only mildly with prompt length, reflecting that they avoid expert transfers on the critical path. ExpertFlow shows the steepest growth and the largest tail amplification. On Qwen3-30B, its average TTFT rises to around 10\,s and the P99 approaches the high teens as the prompt reaches the longest range in the sweep, while \systemname\ stabilizes near the mid-single digits and keeps P99 below ExpertFlow by a wide margin. The separation is larger on Qwen3-80B, where ExpertFlow converges to roughly the mid-40\,s range in average TTFT and close to 90\,s at P99, whereas \systemname\ remains substantially lower and tracks a slower growth trend as tokens increase. Phi-3.5-MoE shows a similar pattern: static Int4 stays near the bottom, ExpertFlow exhibits a rapid rise and a high plateau, and \systemname\ sits in between with a smaller average--tail gap. These results are consistent with longer prompts making prefill less sparse and increasing the volume and frequency of expert movement; bounding transition interference and avoiding blocking publication keeps \systemname's latency growth more gradual as token count increases.

\section{Related Work}

Efficient neural inference under hard device budgets is a recurring systems concern across domains, from resource-constrained perception models for autonomous driving~\cite{zhang2024yoloppa} to large MoE language models whose expert weights exceed single-GPU HBM. We organize prior work by the control axis each line exposes at serving time: where expert weights reside and when they move, versus how precision is assigned and whether that assignment can change online. Our contribution is the runtime execution contract that makes online precision residency budget-safe and non-blocking on a single GPU.

\subsection{MoE Serving Systems and Expert Offloading}
MoE serving systems aim to reduce dispatch overheads, improve expert parallelism, and manage data movement across the GPU, CPU, and storage hierarchy. DeepSpeed-MoE~\cite{rajbhandari2022deepspeed}, Tutel~\cite{hwang2023tutel}, and MegaBlocks~\cite{gale2023megablocks} focus on efficient routing and grouped expert execution, improving kernel utilization and reducing communication and launch overheads. These frameworks largely assume that expert parameters are available in GPU memory, and their primary lever is to restructure computation and scheduling around sparse activation.

A separate line of work treats GPU memory as the bottleneck and explicitly offloads experts to slower tiers. MoE-Infinity~\cite{xue2024moe} characterizes activation locality and uses tracing to guide expert caching and offloading. ProMoE~\cite{song2024promoe} uses proactive caching to reduce cache misses by predicting future expert usage and prefetching ahead of demand. ExpertFlow~\cite{shen2025expertflow} studies cache-aware routing and adaptive scheduling to coordinate prefetching and memory usage across layers. SwapMoE~\cite{kong2024swapmoe} reduces memory footprint by maintaining a small dynamic set of virtual experts and remapping them to physical experts, relying on locality to amortize swapping costs. Fine-grained offloading further refines this trade-off by extracting more detailed activation patterns to guide prefetch and caching decisions~\cite{yu2025fmoe}. Pre-gated MoE changes the algorithmic interface by pre-gating experts to reduce activation volatility and ease system-level scheduling~\cite{hwang2024pre}. These systems primarily answer where expert weights should reside and when to move them, treating weights as fixed-precision objects during serving.

Recent work begins to explore mixed precision in the presence of offloading. HOBBIT proposes a mixed-precision expert offloading design that replaces less critical cache-miss experts with lower-precision versions to reduce loading latency under memory pressure~\cite{tang2024hobbit}. Our work differs in the execution contract and the decision boundary. We adopt stable handles to decouple version visibility from transfers, and we cast precision selection as an online, budget-constrained problem that is driven by routing dynamics. In this view, offloading and caching remain relevant mechanisms, but precision becomes an additional control axis for satisfying a strict HBM budget while bounding interference from background transitions.

\subsection{MoE Quantization and Dynamic Precision Control}
Post-training quantization has become a standard tool for reducing LLM memory footprint and improving throughput. Methods such as SmoothQuant~\cite{xiao2023smoothquant} reduce activation outliers to enable efficient low-bit inference, while weight-only schemes such as GPTQ~\cite{frantar2022gptq}, AWQ~\cite{lin2024awq}, and SpQR~\cite{dettmers2023spqr} improve accuracy at low bitwidths through second-order approximations, activation-aware calibration, or sparse-quantized representations. QuaRot further shows that fixed rotations can remove activation outliers and enable end-to-end low-bit inference, including the KV cache~\cite{ashkboos2024quarot}. These approaches are effective for dense Transformers, but they typically produce a static quantized model, with a fixed precision configuration that does not change during serving.

Dynamic quantization and adaptive inference adjust execution in response to runtime signals such as layer sensitivity, hardware feedback, or input difficulty. HAQ~\cite{wang2019haq} uses hardware-aware search to choose quantization configurations, while AdaQuant~\cite{hubara2021accurate} and FlexRound~\cite{lee2024flexround} improve PTQ by optimizing rounding and calibration to preserve accuracy. These techniques generally operate at the granularity of layers or blocks and are designed for dense networks where all parameters participate in every token. MoE introduces a different structure: expert activations are sparse, heavy-tailed, and sensitive to workload shifts, and the serving system must enforce a hard HBM budget while avoiding blocking transitions.

Several recent papers target quantization specifically for MoE by exploiting expert heterogeneity. MxMoE considers expert sensitivity and activation dynamics when deriving mixed-precision configurations and couples quantization with kernel generation for mixed-precision grouped GEMMs~\cite{duanmu2025mxmoe}. MoPEQ studies mixed-precision expert quantization and assigns different bitwidths to experts based on activation frequency and sensitivity measures~\cite{chittyvenkata2025mopeq}. These methods primarily derive a static mixed-precision assignment or require offline profiling that is not tied to online serving dynamics. In contrast, our focus is on online precision residency under strict device-memory constraints. We rely on router traces to update a budget-feasible high-precision resident set, and we provide a non-blocking realization with window-level pinning, deterministic memory management, and bounded-interference transitions. This execution contract is central to serving, where precision changes must not introduce stalls or destabilize tail latency under concurrent batching and shifting workloads.

\section{Conclusion}

We presented \systemname\ for single-GPU MoE serving under tight HBM budgets, framing the problem as online, budget-constrained precision allocation driven by runtime routing. \systemname\ concentrates high precision on long-horizon hot experts while keeping a low-precision fallback, enforces budget feasibility by construction, and updates residency through asynchronous promotions and demotions so inference proceeds on stable expert versions. Across workloads with skewed and shifting routing, this design improves the quality--memory tradeoff over static PTQ and reduces the waiting latency that limits offloading and prefetching under dense activation.

\nocite{langley00}

\bibliographystyle{ACM-Reference-Format}
\bibliography{references}

\end{document}